# An Extension of the Iterated Moving Average

Edward Valachovic[1]
[1]Department of Epidemiology and Biostatistics, College of Integrated Health Sciences, University at Albany, State University of New York, One University Place, Rensselaer, NY 12144

**Abstract**
This work introduces an extension of the iterated moving average filter, called the Extended Kolmogorov-Zurbenko (EKZ) filter for time series and spatio-temporal analysis. The iterated application of a central simple moving average (SMA) filter, also known as a Kolmogorov-Zurbenko (KZ) filter, is a low-pass filter defined by the length of the moving average window and the number of iterations. These two arguments determine the filter properties such as the energy transfer function and cut-off frequency. However, the existing KZ filter is only defined for positive odd integer widow lengths. Therefore, for any finite time series dataset there is only a relatively small selection of possible window lengths, determined by the length of the dataset, with which to apply a KZ filter. This inflexibility impedes use of KZ filters for a wide variety of applications such as time series component separation, filtration, signal reconstruction, energy transfer function design, modeling, and forecasting. The proposed EKZ filter extends the KZ and SMA filters by permitting a widened range of argument selection for the filter window length providing the choice of an infinite number of filters that may be applied to a dataset, affording enhanced control over the filter characteristics and greater practical application. Simulations and real data application examples are provided.

**Key Words:** Time Series, Spatio-Temporal Analysis, Extended Iterated Moving Average, Kolmogorov-Zurbenko, Low-pass Filter.

## 1. Introduction

In time series data, successive data points, or observations, are ordered in time and are often composed of multiple contributing factors such as trends, periodic components and random variation. The moving average filter is a useful tool in time series analysis that can help smooth shorter-term variation and benefit the study of longer-term patterns and trends free from the interference of obscuring fluctuations. More on time series analysis and moving averages including definitions, notation, and examples can be found in Wei (1990) and Shumway and Stoffer (2017).[1,2] While other moving average designs are possible, a common moving average design is a central simple moving average (SMA). The SMA filter, with notation $SMA_m$, replaces each observation in a time series with an average of a chosen odd positive integer number, $m$ or the filter window length, of sequential observations from the time series centered at the observation being replaced. While other filter designs can assign different weights to the values being averaged, increasing or decreasing their influence, the SMA assigns a weight of one to each observation, so that each has the same influence on the final average.

With the design described above, an $SMA_m$ helps to smooth short-term fluctuations that occur with a period of $m$ and shorter. A few simple examples illustrate the practical applications of an $SMA_m$. Consider for instance, a time series dataset consisting of any purely periodic pattern with a period of *m*, or equivalently a frequency or reciprocal of the period, *1/m*, such as a sinusoidal wave. With the exception the first and last *(m-1)/2* observations of the time series, where the SMA filter does not have enough observations to be fully calculated, the $SMA_m$ applied to this time series will produce a constant value, the overall mean, everywhere else. Since *m* equals the period of the sinusoidal wave, the $SMA_m$ calculations will be composed of a balanced set of values about the overall mean. Relating observations made along a sinusoid to coordinates of a unit circle, the same idea can be thought of as having any odd number, *m*, set of points equally dividing the unit circle, where the average of the coordinates of those *m* points completing the unit circle are equal to zero. Clearly, an $SMA_m$ filter is useful to eliminate components with a period of *m*, or frequency *1/m*. Using the same reasoning, the $SMA_m$ filter is also able to completely suppress components at the harmonics, or positive integer multiple of the fundamental frequency *1/m*. Consider now the same periodic time series, but with a single random fluctuation or interruption to the pattern at one time point. Through the process of averaging, the same $SMA_m$ will smooth that fluctuation but not completely suppress its effect upon the surrounding nearest *m* observation values with which it is averaged. This example illustrates one potential drawback of the $SMA_m$ filter since it suppresses or attenuates but does not necessarily eliminate noise or variation that has a period shorter than *m*.

An extension of the $SMA_m$ filter was introduced by Zurbenko (1986) which improved upon the filter design and more completely suppressed variation with a period less than *m*, or frequency greater than *1/m*.[3] The extension, called the Kolmogorov-Zurbenko (KZ) filter, is the iteration of the SMA filter. With notation $KZ_{m,k}$ the KZ filter has the two arguments *m*, the positive odd integer filter window length, and *k*, the positive integer number of iterations. Therefore, the $KZ_{m,k}$ filter is equal to the $SMA_m$ filter when $k = 1$. With definitions and properties provided in Yang and Zurbenko (2010), the $KZ_{m,k}$ filters are a class of low-pass filters, completely attenuating variation within a time series at a frequency of *1/m* as well as its harmonics.[4] However, when $k > 1$, the $KZ_{m,k}$ filter more fully suppresses all variation that has a period shorter the *m*, and the level of attenuation can be controlled by the choice of argument *k*.

Yang and Zurbenko (2010) provides the energy transfer function for the KZ filter, which is a mapping that describes how input frequencies are transferred to outputs.[4] More on the energy transfer function can be found in Shumway and Stoffer (2017) and Zurbenko (1986).[2,3] The energy transfer function is useful to show the effect of a KZ filter on a time series, and how with only a few iterations, this class of filters, strongly attenuates signals of frequency 1/*m* and higher while passing lower frequencies. Figure 1 illustrates examples of KZ filters, all with a fixed argument of $m = 7$ but paired with different choices for the argument of the number of iterations. The energy transfer functions for $k = 1$ is in magenta which is equivalent to an $SMA_{m=7}$, for $k = 2$ in red, $k = 3$ in yellow, $k = 4$ in green, $k = 5$ in cyan, and $k = 6$ in blue. The figure on the left displays the energy transfer functions and shows that the attenuation is complete at frequency *1/m* = 1/7 and its harmonics 2/7 and 3/7, but the $KZ_{m=7,k=1}$ or equivalently $SMA_{m=7}$ in magenta only modestly suppresses other frequencies greater than *1/m*. The advantage of the KZ filters over SMA filters is that higher iterations improve this attenuation. At two iterations in red, the $KZ_{m=7,k=2}$ more completely suppresses these other frequencies, but there is still visible evidence of leakage through the filter at higher frequencies. At $k = 3$ and above, in the figure to the left, the energy transfer function does not appear to be visibly different from zero at frequencies greater than *1/m*. However, in the figure to the right the natural logarithm of the energy transfer functions reveals that this is not quite the case. As the number of iterations increases the KZ filters both completely attenuate at frequency *1/m* and its harmonics, and the degree of attenuation at frequencies greater than *1/m* increases.

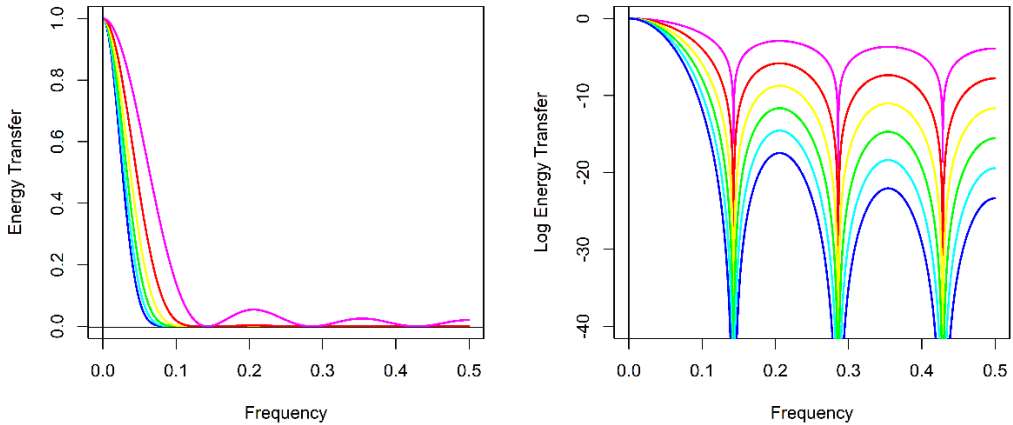

**Figure 1**: The energy transfer function to the left and log energy transfer function to the right for Kolmogorov-Zurbenko filters with a fixed argument of $m = 7$ paired with $k = 1$ in magenta, $k = 2$ in red, $k = 3$ in yellow, $k = 4$ in green, $k = 5$ in cyan, and $k = 6$ in blue.

The KZ filter is a notable improvement upon the SMA filter, overcoming the inability of the SMA filter to adequately attenuate frequencies greater than *1/m* which limits its use as a low-pass filter. Furthermore, the KZ filter provides direct control through one of its functional arguments, the number of iterations, *k*, that adjusts the degree of attenuation in this range of frequencies allowing more precise design of the filter for different applications. KZ filters and their extensions are useful to separate portions of the frequency domain to exclude interfering frequencies. These filters have a history of use in a variety of fields such as the environmental sciences, meteorology, and climatology in Wise and Comrie (2005) and Gao et al. (2021).[5,6] KZ filters featured in studying air quality in Kang et al. (2013) and Sezen et al. (2023).[7,8] Zhang, Ma, and Kim (2018) used KZ filters to study pollution and De Jongh, Verhoest and De Troch (2006) used them to research precipitation patters.[9,10] Valachovic and Zurbenko uses KZ filters for frequency separation to identify hidden periodically correlated components in skin cancer time series data and perform multivariate analysis on these component factors (2017).[11] Recently, they were used by Valachovic to design a new method called the Variable Bandpass Periodic Block Bootstrap to address a weakness in prior block bootstrapping methods for periodically correlated time series.[12] Many of these examples highlight the use of KZ filters to smooth data, reduce random variation, interpolate missing observations, and specifically separate and filter portions of the frequency domain prior to analysis.

However, one design shortcoming of the KZ filter is its constraint in the selection of filter window length. With a design that is restricted to positive odd integers for the argument *m*, it is severely limiting in the choice of which period or corresponding frequency can be completely suppressed. With choice restricted to positive odd integers, there are continuous interval gaps of possible periodic signal periods that can not be used. Figure 2 shows this shortcoming in the energy transfer functions for KZ filters with a fixed argument of $k = 1$ paired with $m = 3$ in red, $m = 5$ in green, $m = 7$ in blue, $m = 9$ in dashed red, $m = 11$ in dashed green, $m = 13$ in dashed blue. Clearly, as $m$ increases, the energy transfer functions crop near zero along with the frequency $1/m$ which is completely attenuated. As a result, the design options of KZ filters are sparse at relatively small values for the argument $m$ than at higher values of $m$. Seen in Figure 2, there are only four choices for the filter window length in KZ filters between the frequencies of 0.1 and 0.5.

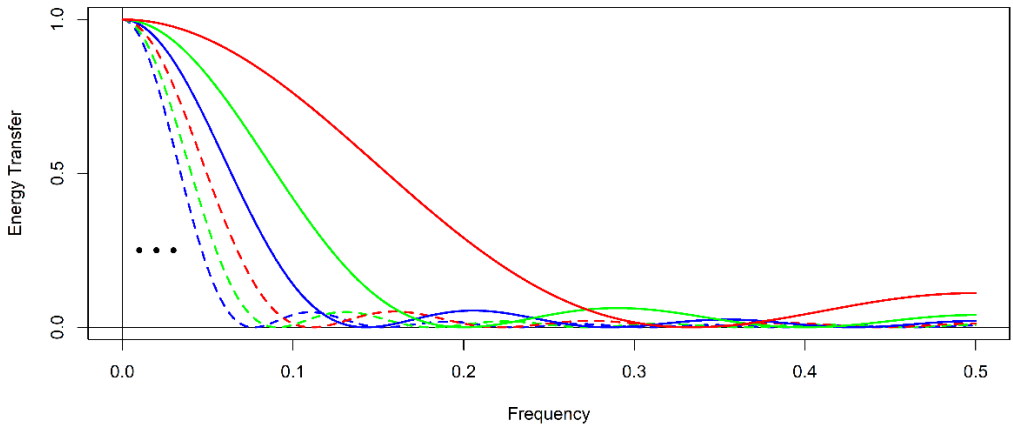

**Figure 2**: The energy transfer function for Kolmogorov-Zurbenko filters with a fixed argument of $k = 1$ paired with $m = 3$ in red, $m = 5$ in green, $m = 7$ in blue, $m = 9$ in dashed red, $m = 11$ in dashed green, $m = 13$ in dashed blue. The ellipsis shows that there are additional KZ energy transfer functions in this region as $m$ increases.

This is very limiting when designing KZ filters for signal separation and reconstruction. For instance, it may be the case that a component signal with a positive even integer period is the target for filtration. While it is possible to approximate a filter for a signal with an even period using a KZ filter designed with an adjacent odd period, the filtration will be incomplete. Furthermore, as Figure 2 illustrated, the approximation is increasingly poor for smaller values of $m$. For other examples, even when a signal has an odd integer period, such as a weekly or 7-day period, or yearly 365-day period, if the goal is separating harmonics of the fundamental frequency, those harmonics will have non-integer periods which can not be designed into the KZ filter window length.

The Extended Kolmogorov-Zurbenko (EKZ) filter, introduced here, is designed to solve this limitation in the KZ filter. The EKZ is an iteration of a generalization of a simple moving average filter so that it permits any positive real valued window length and is formally defined in the next section. In summary, the EKZ creates the equivalent of a centered and equally weighted moving average, somewhat counterintuitively, using a specially designed non-equally weighted moving average. To accomplish this, it treats a time series, with its original unit time measure, as if it was recorded over intervals of smaller time units, or even continuously. Since, any positive real valued number greater than one, $m_r$, can be expressed as the sum of the closest positive odd integer less than that real number, $m_o$, and the difference, $m_d$, which must be less than two. Therefore, any positive real valued window length, $m_r$, can be used to create an extended central simple moving average formed from the original values of the original time series, with weights 1 for the nearest $m_o$ observations surrounding the central time point, and weights equal to half of $m_d$ for the next farther observations before and after. With this design, the EKZ is identical to a KZ filter if it had it been applied to the time series were measured in finer time units than original. Therefore, the EKZ is an extension of the KZ filter as an iterated simple moving averaged, and it retains many properties of the KZ filter, while permitting a continuous choice of filter window length $m_r$.

## 2. Methods

### 2.1 The Iterated Moving Average

The original Kolmogorov-Zurbenko (KZ) filter is the iteration of a central simple central moving (SMA) average defined in Zurbenko (1986).[3] It is a filter with two arguments $m$, and $k$, and notation $KZ_{m,k}$. The argument $m$, a positive odd integer, is the length of the SMA filter window, and the argument $k$, a positive integer, is the number of iterations of the SMA. Therefore, the $SMA_m$ is a special case of the $KZ_{m,k}$ filter, when $k=1$. KZ filters are a type of low-pass filter that completely attenuates signals at frequency $1/m$ and its integer multiples or harmonics, and strongly attenuates all signals of frequency $1/m$ and higher while passing lower frequencies. The effect of a $KZ_{m,k}$ filter is smoothing the time series regarding the variation that occurs with period $m$ or shorter. The equations below follow the notation found in Zurbenko (1986) and Yang and Zurbenko (2010).[3,4] Applied to a time series process $\{X(t): t \in \mathbb{Z}\}$, a KZ filter of $m$ data-points and $k$ iterations, where $t$ is time ordering the data-points, $u$ are the time steps defining the symmetric centered filter window, and $a_u^{m,k}$ are coefficient weights, is defined as follows:

$$KZ_{m,k}(X(t)) = \sum_{u=-\frac{k(m-1)}{2}}^{\frac{k(m-1)}{2}} \frac{a_u^{m,k}}{m^k} X(t+u) \qquad (2.1.1)$$

The coefficients $a_u^{m,k}$ can be conveniently found as the polynomial coefficients obtained from an expansion of the following polynomial (arbitrarily in $z$):

$$\sum_{r=0}^{k(m-1)} z^r a_{r-\frac{k(m-1)}{2}}^{m,k} = (1 + z + \cdots + z^{m-1})^k \qquad (2.1.2)$$

The coefficients are obtained by the convolution of k uniform discrete distributions on the interval $\left[-\frac{m-1}{2}, \frac{m-1}{2}\right]$ where $m$ is an odd integer. Therefore, each coefficient forms a tapering window which has finite support. For some examples calculating the coefficients, a $KZ_{m=3,k=1}$ filter would have an ordered sequence of $a_u^{3,1}$ coefficients $\{1,1,1\}$ like an $SMA_3$, and a $KZ_{m=3,k=2}$ filter would have ordered sequence of $a_u^{3,2}$ coefficients $\{1,2,3,2,1\}$.

The KZ filter can be applied with statistical software in an iterated form using R Studio software packages such as KZA detailed in Close and Zurbenko (2013).[13] The iterated form of the Kolmogorov-Zurbenko filter can be produced according to the following algorithm found in Yang and Zurbenko (2010) along with the following equations.[4] In the iterated form of the $KZ_{m,k}$ filter each step is an application of a $KZ_{m,k=1}$ filter, or equivalently an $SMA_m$ moving average filter of length $m$, to the prior result thus making all $a_u^{m,1}$ coefficients equal to one in each step. The iterated form of the $KZ_{m,k}$ is provided in the following equations:

$$KZ_{m,1}(X(t)) = \sum_{u=-\frac{m-1}{2}}^{\frac{m-1}{2}} \frac{a_u^{m,1}}{m^1} X(t+u) = \frac{1}{m} \sum_{u=-\frac{m-1}{2}}^{\frac{m-1}{2}} X(t+u) \qquad (2.1.3)$$

$$KZ_{m,2}(X(t)) = \frac{1}{m} \sum_{u=-\frac{(m-1)}{2}}^{\frac{(m-1)}{2}} KZ_{m,1}(X(t+u)) \qquad (2.1.4)$$

$$\vdots$$

$$KZ_{m,k}(X(t)) = \frac{1}{m} \sum_{u=-\frac{(m-1)}{2}}^{\frac{(m-1)}{2}} KZ_{m,k-1}(X(t+u)) \qquad (2.1.5)$$

Additionally, Zurbenko (1986) and Yang and Zurbenko (2010) introduce the energy transfer function for the KZ filter, which is the square of the linear mapping that describes how input frequencies are transferred to outputs and as such is symmetric about zero.[3,4] The energy transfer function is useful to show the effect of a KZ filter on a time series, and how with only a few iterations, this class of filters, strongly attenuates signals of frequency $1/m$ and higher while passing lower frequencies. The energy transfer function of the KZ filter at each frequency $\lambda$ is seen in the following equation:

$$|B(\lambda)|^2 = \left(\frac{\sin(\pi m \lambda)}{m \sin(\pi \lambda)}\right)^{2k} \qquad (2.1.6)$$

The KZ energy transfer function is also useful to define the cut-off frequency which is a limit or boundary at which the energy transferred through a filter is generally suppressed or diminished rather than allowed to pass through. A cut-off frequency where output power is half that of the input, called the half power point, for the KZ filter energy transfer function is provided by Zurbenko (1986) and Yang and Zurbenko (2010) below.[3,4]

$$\lambda_0 \approx \frac{\sqrt{6}}{\pi} \sqrt{\frac{1 - \left(\frac{1}{2}\right)^{\frac{1}{2k}}}{m^2 - \left(\frac{1}{2}\right)^{\frac{1}{2k}}}} \qquad (2.1.7)$$

**2.2 The Extended Iterated Moving Average**
The proposed extension to the iterated moving average introduced here, equivalent to an extension of the KZ filter which itself is an extension of the SMA filter, is called the

Extended Kolmogorov-Zurbenko (EKZ) filter. The design of the EKZ arises from an aim to define a central SMA or KZ filter over both ever shorter and finer time units. By design a central SMA filter averages all data observations with equal weight in the filter window length. Similarly, the KZ filter balances the number of observations before, and after which centralizes the filter and results in the limitation to only positive odd integer window lengths. Rather than equally weighting observations, in order to define a central equally weighted moving average for numbers other than positive odd integer window lengths, the proposed EKZ creates an extension of the KZ filter by iterating a specially, and unequally, weighted central moving average filter on a positive odd integer window length. In effect, the EKZ filter creates moving averages that are equal to what they would be with KZ filters, were the data recorded finely enough so that the desired window length was a positive odd integer. This permits the EKZ to be applied as if it were an SMA applied to any window length, including those that are not positive odd integer lengths. Therefore, it is a filter with notation $EKZ_{m_r,k}$ and with two arguments $m_r$, and $k$, where the argument $m_r$ is any positive real number greater than one for the filter window length. The argument $k$ remains a positive integer for the number of iterations. As noted previously, where a $KZ_{m,k=1}$ filter is equivalent to an $SMA_m$ moving average filter of length $m$, by extension, an $EKZ_{m_r,k=1}$ filter is equivalent to an extension of an SMA filter for filter length $m_r$, which can be called an $ESMA_{m_r}$.

To define the EKZ filter, let $\{X(t): t \in \mathbb{Z}\}$ be a time series where $t$ is time. For an $EKZ_{m_r,k}$ filter of a positive real valued filter window length of $m_r > 1$ time points and positive integer $k$ iterations, define $m_r = m_o + m_d$ time points where $m_o$ is the nearest positive odd integer less than $m_r$, and $m_d$ is the difference equal to $m_r - m_o$. Clearly, it must be that $0 \leq m_d < 2$. Then an $EKZ_{m_r,k}$ filter of $m_r$ time points and $k$ iterations, where $u$ are the steps defining the symmetric filter window, and $a_u^{m,k}$ are coefficient weights, is defined as follows:

$$EKZ_{m_r,k}(X(t)) = \frac{1}{(m_r)^k} \sum_{u=-\frac{k(m_o+1)}{2}}^{\frac{k(m_o+1)}{2}} a_u^{m_r,k} X(t+u) \qquad (2.2.1)$$

The coefficients $a_u^{m,k}$ can be conveniently found as the polynomial coefficients obtained from an expansion of the following polynomial (arbitrarily in $z$):

$$\sum_{h=0}^{k(m_o+1)} z^h a_{h-\frac{k(m_o+1)}{2}}^{m_r,k} = \left(\frac{m_d}{2} + z + \cdots + z^{m_o} + \frac{m_d}{2} z^{m_o+1}\right)^k \qquad (2.2.2)$$

The coefficients are obtained by the convolution of k uniform discrete distributions on the interval $\left[-\frac{m_o+1}{2}, \frac{m_o+1}{2}\right]$ where $m_o$ is an odd integer, and each coefficient forms a tapering window which has finite support. The difference in the EKZ filter from the KZ filter is that the first and last coefficients in the polynomial are $\frac{m_d}{2}$ rather than one. For some examples calculating the coefficients, an $EKZ_{m_r=1.5,k=1}$ filter would have an ordered sequence of $a_u^{1.5,1}$ coefficients $\{0.25, 1, 0.25\}$. An $EKZ_{m_r=1.5,k=2}$ filter would have an ordered sequence of $a_u^{1.5,2}$ coefficients $\{0.25, 1.25, 1.5, 1.25, 0.25\}$. Lastly, a $EKZ_{m_r=\pi,k=1}$ would have ordered sequence of $a_u^{\pi,1}$ coefficients $\left\{\frac{\pi-3}{2}, 1, 1, 1, \frac{\pi-3}{2}\right\}$.

In the iterated form of the $EKZ_{m_r,k}$ filter each step is an application of an $EKZ_{m_r,1}$ filter, or equivalently an $ESMA_{m_r}$, to the prior result. In this form, since $k=1$ in each step, all $a_u^{m_r,k}$ coefficients are equal to one in each step except for the first and last coefficients provided $m_d \neq 0$. The iterated form of the $EKZ_{m_r,k}$ is provided in the following equations:

$$EKZ_{m_r,1}(X(t)) = \frac{\frac{1}{2}m_d}{m_r} X\left(t - \frac{m_o+1}{2}\right) + \left[\sum_{u=-\frac{m_o-1}{2}}^{\frac{m_o-1}{2}} \frac{1}{m_r} X(t+u)\right] + \frac{\frac{1}{2}m_d}{m_r} X\left(t + \frac{m_o+1}{2}\right) \quad (2.2.3)$$

$$EKZ_{m_r,2}(X(t)) = \frac{\frac{1}{2}m_d}{m_r} EKZ_{m_r,1}\left(X\left(t - \frac{m_o+1}{2}\right)\right) + \left[\sum_{u=-\frac{m_o-1}{2}}^{\frac{m_o-1}{2}} \frac{1}{m_r} EKZ_{m_r,1}(X(t+u))\right] + \frac{\frac{1}{2}m_d}{m_r} EKZ_{m_r,1}\left(X\left(t + \frac{m_o+1}{2}\right)\right) \quad (2.2.4)$$

$$\vdots$$

$$EKZ_{m_r,k}(X(t)) = \frac{\frac{1}{2}m_d}{m_r} EKZ_{m_r,k-1}\left(X\left(t - \frac{m_o+1}{2}\right)\right) + \left[\sum_{u=-\frac{m_o-1}{2}}^{\frac{m_o-1}{2}} \frac{a_u^{m,1}}{m_r} EKZ_{m,k-1}(X(t+u))\right] + \frac{\frac{1}{2}m_d}{m_r} EKZ_{m,k-1}\left(X\left(t + \frac{m_o+1}{2}\right)\right) \quad (2.2.5)$$

EKZ filters inherit or are approximated by many of the properties of KZ filters, because KZ filters are a special subset of EKZ filters when argument $m_r = m$ as a positive odd integer. So just like the KZ filter, the $EKZ_{m_r,k}$ filter smooths variation in a time series that occurs with period $m_r$ or shorter and smooths more completely with greater iterations. To better understand the properties, three cases must be considered: when the window length is a positive odd integer, when it is a positive even integer, and when it is any other real number greater than one. EKZ filters where $m_r = m$, a positive odd integer, mathematically simplify to KZ filters, which completely attenuates signals at frequency $1/m$ and its multiples, or harmonics, and strongly attenuates all signals of frequency $1/m$ and higher while passing lower frequencies.[3,4] This was described earlier with an example of a sinusoidal wave of period $m$. With the exception of the beginning and ending observations of the time series, where the filter is not completely implemented, the $EKZ_{m_r,k}$ just like the $KZ_{m,k}$ or $SMA_m$ filter will produce a constant value, the overall mean, everywhere else. Again, since $m$ equals the period of the sinusoidal wave, the $SMA_m$ calculations will be composed of a balanced set of values about the overall mean. Equivalently, the same idea can be thought of as having any odd number $m$, set of points equally dividing the unit circle, where the average of the coordinates of those $m$ points completing the unit circle are equal to zero.

When $m_r$ is a positive even integer, a case that the KZ and SMA filter can not accommodate, the EKZ filter also completely attenuates signals at frequency $1/m_r$ and harmonics, and strongly attenuates all signals of frequency $1/m_r$ and higher while passing lower frequencies. For a sinusoidal wave of period $m_r$, with the exception the beginning and ending observations of the time series, where the filter is not completely applied, the $EKZ_{m_r,k}$ will produce a constant value, the overall mean, everywhere else. Since $m_r$ equals the period of the sinusoidal wave, the calculations of the EKZ will still be composed of a balanced set of values about the overall mean. However, since $m_r$ is even, each calculated average will contain $m_r + 1 = m_o + 2$ values, including the central most $m_o$ values, and the two next closest values, one preceding and one following but each weighted by half. Equivalently, this time the idea can be thought of as having any even number, $m_r$, set of points equally dividing the unit circle. The EKZ averages $m_r + 1 = m_o + 2$ values, weighting $m_o$ values by one, and repeating the last value in the calculation twice, but at the same time weighting the repeated values by half. This is equivalent to an average of the $m_r$ set of points and that average always equals zero.

However, when $m_r$ is any other real number greater than one, the properties of the EKZ filter are only approximated by the properties of the KZ filter. When $m_r$ is any other real number greater than one, the EKZ filter very strongly attenuates, but not completely attenuates, signals at frequency $1/m_r$ and harmonics, while it still passes lower frequencies and strongly attenuates all signals of frequency $1/m_r$ and higher. In the case of a sinusoidal wave of period $m_r$, no integer fully divides $m_r$, and the calculations of the EKZ will be close to but no longer be composed of a balanced set of values about the overall mean. The reason the EKZ calculations will be close to but not exactly equal to the overall mean can again be seen by considering the unit circle. In this case, the unit circle is divisible by $m_r$ points but not by an integer due to the fixed size of the unit measure of the time series. So, $m_o$ points and corresponding equal size intervals do not completely divide the unit circle or one cycle of period $m_r$, and $m_o + 2$ intervals are more than the unit circle. Therefore, the EKZ calculation averaging a set of $m_o$ points weighted by one, and the two next closest points by half of $m_d = m_r - m_o$, will be close to but not exactly equal to zero. Since not all EKZ averages are equal, variation at a frequency of $1/m_r$ as well as it's harmonics are strongly but not completely suppressed or eliminated.

Therefore, the energy transfer function inherited from the KZ filter is best considered as only approximate for the EKZ filter at each frequency $\lambda$, specifically the same when $m_r$ is a positive odd greater than one, and approximate for all other values greater than one, and is seen in the following equation:

$$|B(\lambda)|^2 = \left(\frac{\sin(\pi m_r \lambda)}{m_r \sin(\pi \lambda)}\right)^{2k} \tag{2.2.6}$$

Likewise, the cut-off frequency where output power is half that of the input, or the half power point, for the EKZ filter energy transfer function is approximated by the equation provided below:

$$\lambda_0 \approx \frac{\sqrt{6}}{\pi}\sqrt{\frac{1-\left(\frac{1}{2}\right)^{\frac{1}{2k}}}{(m_r)^2 - \left(\frac{1}{2}\right)^{\frac{1}{2k}}}} \tag{2.2.7}$$

Figure 3 shows the energy transfer function for multiple EKZ filters over a small range of filter window lengths, many of which are not possible with KZ or SMA filters. To demonstrate flexibility, this figure shows EKZ filters with the number of iterations fixed at $k = 1$, and the moving average window length ranging from $m_r = 1$ to $m_r = 7$ by one-unit intervals. Between each interval, values of $m_r$ are used to illustrate the designs of filters possible. Of course, this is only illustrative of the range of EKZ filters possible, of which there are an infinite number for any real number within this range. For comparison, there are only three KZ filters, or equivalently SMA filters, in the same range where $m = 3, 5,$ or $7$, and whose energy transfer functions are identified by the black dashed lines in the figure. Finally, more filters are possible as $m_r$ increases above 7, and the energy transfer function becomes more compact around the zero-frequency indicated by the ellipsis in the figure.

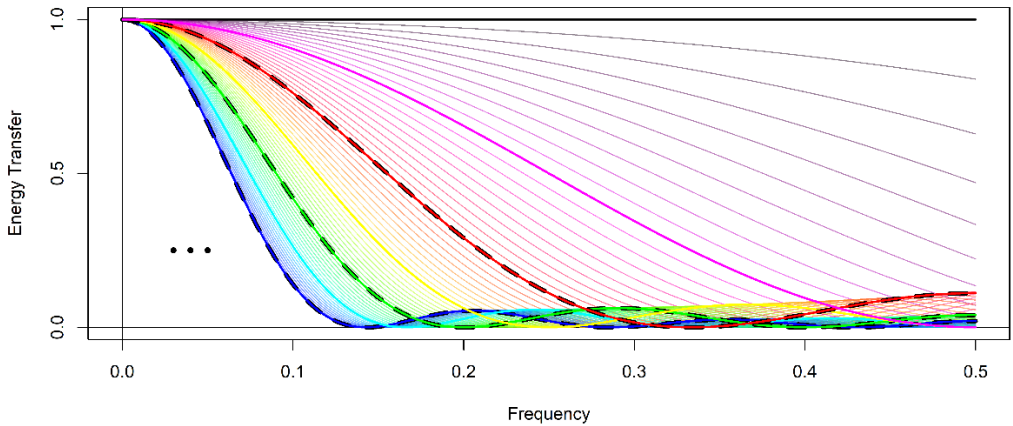

**Figure 3**: The energy transfer function for Extended Kolmogorov-Zurbenko filters with arguments $k = 1$ and $m_r = 1$ in black, $m_r = 2$ in magenta, $m_r = 3$ in red, $m_r = 4$ in yellow, $m_r = 5$ in green, $m_r = 6$ in cyan, and $m_r = 7$ in blue, in addition to $m_r$ within each created interval. The lines with black dashes are the only possible energy transfer functions for KZ filters in the same range of argument $m$ at $m = 3$ in red, $m = 5$ in green, and $m = 7$ in blue.

### 3. Simulations

To illustrate the use of the EKZ filter proposed here as well as the results obtain from their application, this work simulates a time series and applies EKZ, or extended iterated moving average, filters with arguments that are not possible with KZ or SMA filters. Periodograms, which are described in Wei (1990) and Shumway and Stoffer (2017), and illustrate sample variance at each frequency within a time series, provide a visual tool to observe the varied effects of different EKZ and KZ filters upon the original time series.[1,2]

Analysis is performed in R version 4.1.1 (2013) statistical software using the KZ function in the KZA package, see Close and Zurbenko (2013) for more detail, with datasets as a time series measured on an ordered interval dimension, in this case time.[13,14] The simulated time series is constructed with 100,000-time units. First, a time series of random variation or white noise is generated with independent draws from a standard normal distribution with mean zero and standard deviation one. The periodogram of this white noise time series data is seen in Figure 4, plotted using R in black, and exhibits random low fluctuating levels of energy across all frequencies which is characteristic of white noise. The periodogram of an $EKZ_{m_r=2,k=1}$ filter applied to this time series in magenta, exhibits the successful smoothing and complete suppression of variation at a frequency of $1/m_r = 1/2$. This matches the energy transfer function of an $EKZ_{m_r=2,k=1}$ filter plotted as the magenta curve with corresponding scale on the right. The periodogram of a second iteration of the $EKZ_{m_r=2,k=1}$ filter, equivalent to an $EKZ_{m_r=2,k=2}$ of the white noise time series, is seen in dark magenta. Note that the $EKZ_{m_r=2,k=2}$ still completely suppresses variation at a frequency of $1/m_r = 1/2$. This matches the energy transfer function of an $EKZ_{m_r=2,k=2}$ filter plotted as the dark magenta curve. For comparison, the closest possible KZ, or equivalent SMA, filter to the $EKZ_{m_r=2,k=1}$ has the energy transfer function displayed in red with $m = 2$ and $k = 1$, which would clearly not suppress variation at a frequency of 1/2.

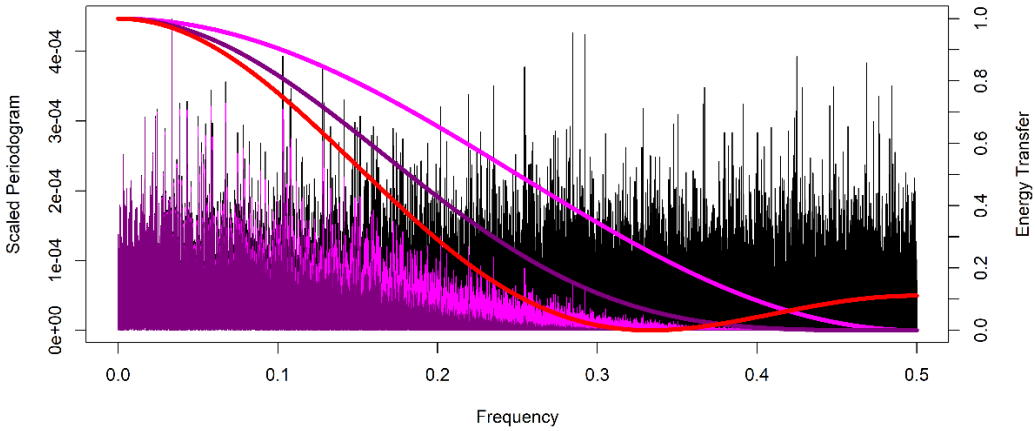

**Figure 4**: The periodogram of a white noise time series in black and the periodograms of that same time series after applying Extended Kolmogorov-Zurbenko filters $EKZ_{m_r=2,k=1}$ in magenta, and $EKZ_{m_r=2,k=2}$ in dark magenta. The energy transfer functions of an $EKZ_{m_r=2,k=1}$ filter is shown as the magenta curve, $EKZ_{m_r=2,k=2}$ in dark magenta, and the closest $EKZ_{m=3,k=1}$, or equivalent $SMA_{m=3}$, filter energy transfer function displayed in red with corresponding scale on the right.

In a second example, an EKZ filter with an example fractional window length is applied to the same white noise time series as the first example. This is not possible with the KZ filter. Due to small values, the log periodogram of the same white noise time series data is seen in Figure 5, plotted in black. The log periodogram of an $EKZ_{m_r=1/0.26,k=1}$ filter applied to this time series in yellow, exhibits the successful smoothing and strong suppression of variation at a frequency of $\frac{1}{m_r} = 0.26$. This matches the approximate energy transfer function of an $EKZ_{m_r=1/0.26,k=1}$ filter plotted as the yellow curve with corresponding scale on the right. For comparison, the closest possible KZ filters $KZ_{m=3,k=1}$ and $KZ_{m=5,k=1}$ have the energy transfer function displayed in red and green respectively.

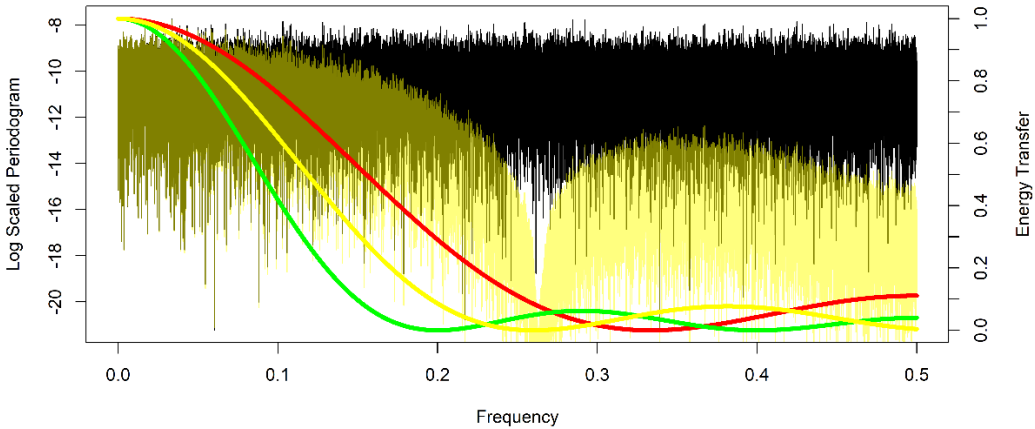

**Figure 5**: The log periodogram of a white noise time series in black and the periodograms of that same time series after applying Extended Kolmogorov-Zurbenko filters $EKZ_{m_r=1/0.26, k=1}$ in yellow. The approximate energy transfer functions of the $EKZ_{m_r=1/0.26, k=1}$ filter is shown as the yellow curve, and the closest possible $KZ_{m=3, k=1}$ displayed in red and $KZ_{m=5, k=1}$ displayed in green with corresponding scale on the right.

## 4. Real Data Applications

The practical use for researchers and potential advantages of the EKZ filter over the KZ filter can be demonstrated in several real data applications. In the first example of a real-world application with an EKZ filter, atmospheric surface pressure records near Seattle, Washington, USA, at 47°46'35.0"N latitude and 122°19'59.0"W longitude, are used from the National Oceanic and Atmospheric Administration Physical Sciences Laboratory data repository NCEP Reanalysis dataset.[15] Data is provided by the NOAA Physical Sciences Laboratory, Boulder, Colorado, USA, from their website at https://psl.noaa.gov/. These records are measured in *Pa*, or Pascals, and are recorded at equal 6-hour intervals four times daily between January 1st, 2010, through December 31st, 2010. The time series has 1,460 observations. Analysis is performed in R version 4.1.1 (2013) statistical software.[13]

The research application may require isolating and eliminating, or alternatively retaining with the combined use of a difference filter, the daily variation exhibited in the time series. The goal here is to apply a filter that completely suppresses variation that occurs over daily periods. While the SMA and KZ filters can set the window length to either *m* = 3 or 5, and this may be sufficient for some purposes, with measurements recorded four times daily completely suppressing daily variation require a moving average with filter window length of four observations. The EKZ filter provides the flexibility to assign the window size to $m_r$ = 4. Figure 6 shows the Seattle surface pressure time series in black and the $EKZ_{m_r=4, k=2}$ filtered time series in yellow. For comparison, the $KZ_{m=3, k=2}$ filtered time series is in red, and the $KZ_{m=5, k=2}$ filtered time series is in green. Figure 6 shows at times noticeable differences between these filters. This result is not surprising given there are significant differences between the energy transfer functions identified in Figure 3, red and green with black dashes for the KZ filter with *m* = 3 and 5 respectively, and yellow for an

EKZ filter with $m_r = 4$ in yellow. With these three filters, $1/m$ is substantially different, as is the case when $m$ is small.

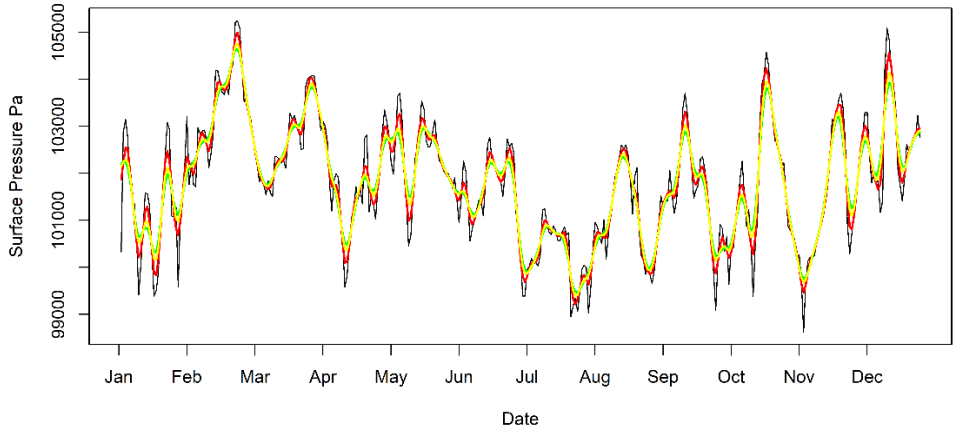

**Figure 6**: The Seattle, Washington surface pressure time series in black and the $EKZ_{m_r=4,k=2}$ filtered time series in yellow, $KZ_{m=3,k=2}$ filtered time series is in red, and the $KZ_{m=5,k=2}$ filtered time series is in green.

In the second example of a practical application with an EKZ filter, daily noon precipitation rate records near New York City Hall, New York, USA, at 40°42'46.0"N latitude and 74°00'21.0"W longitude, are used from the National Oceanic and Atmospheric Administration Physical Sciences Laboratory data repository NCEP Reanalysis dataset.[15] These records are measured in $Kg/m^2/s$ and are recorded between January 1st, 1960, through December 31st, 2019.

For a variety of purposes such as research into linear trends, long term trends, or multiyear patterns, it may be necessary to separate, filter, and eliminate the strong seasonal pattern exhibited in the time series. The goal would be to apply a filter that completely suppresses variation that occurs seasonally, or with a period of one year. This process may typically involve using a simple moving average of 365 days, the $SMA_{m=365}$ or equivalently $KZ_{m=365,k=1}$, or possibly an iterated $KZ_{m=365,k}$ filter with $k>1$ which would strongly suppress all shorter period variation as well. However, depending upon the required precision for the filter this can be problematic for at least two reasons, since the 365-day period only approximates the annual period, and the dataset contains an additional leap year observation every four years, making these filters not as accurate as they could be. A more accurate annual period would be 365.25 days, and for yet greater accuracy a period of 365.256363004 days is closer to the true annual period in days. The SMA and KZ filters can potentially set the window length to either 365 or 367. Since *m* in this example is large compared to the previous example, the difference between the frequency *1/m* for these filters will be small. While any of these may be sufficient for some purposes, they do not have the flexibility of the EKZ filter to assign the window size as arbitrarily close to the annual period as desired. Figure 7 shows the New York City precipitation rate time series in black and the $EKZ_{m_r=365.256363004,k=3}$ filtered time series in red. In practice, for this example, only a very small difference results between the $KZ_{m=365,k=3}$, $KZ_{m=367,k=3}$, and the $EKZ_{m_r=365.256363004, k=3}$ filters, and it is too small to visualize in Figure 7. Still, the difference is measurable between the filter results and the EKZ produces results that more accurately targets suppression of the annual periodic variation from the time series. This example demonstrates the potential use of the greater flexibility of the EKZ filter.

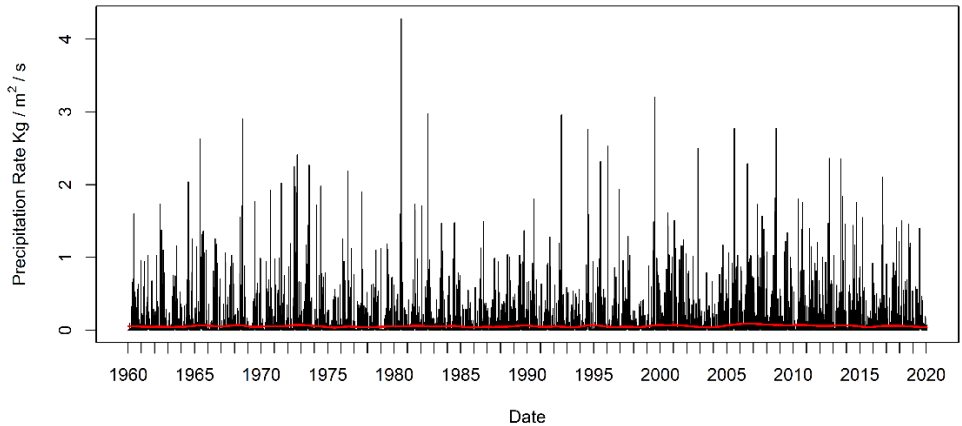

**Figure 7**: The New York City precipitation rate time series in black and the $EKZ_{m_r=365.256363004, k=3}$ filtered time series in red.

## 5. Discussion

The central SMA filter is a useful tool in time series analysis to smooth data, and to completely suppress signals with a period equal to the length of the SMA filter window, as well as it's harmonic frequencies. However, most signals with a period shorter than the window filter length are not completely suppressed, and certain high frequencies have a substantial amount of leakage, or energy that passes through the filter. This makes the SMA a relatively poor low-pass filter. The KZ filter improves upon the SMA and is more effective at attenuating signals with a period shorter than the window filter length by iterating the SMA. The KZ filter creates a better low-pass filter than the SMA, particularly with higher iterations. However, both the KZ and SMA filters are limited because they are only defined for positive odd integer window sizes. This creates a set of filters that has gaps between the frequencies that can be eliminated. This problem is particularly apparent at low positive odd integer values, which account for the largest intervals of the frequency range. The introduction of the EKZ filter in this work helps to fill those gaps, making a much more flexible low-pass filter and overcomes some of the shortcomings of the KZ and SMA filters. The first advantage of the EKZ is that it contains the KZ and SMA filters as special cases, so each of those types of filters and their attributes are achievable. The next advantage of the EKZ is that it permits the use of positive even integer window sizes. This can be done while maintaining the ability of the filter to strongly attenuate signals with a period shorter than the filter window length, while completely eliminating signals with a period equal to, or a multiple of, the filter window length. Just by permitting positive even integer window sizes, the EKZ approximately doubles the number of filter designs that are possible compared to KZ filters while maintaining the same attributes. Finally, when the filter window length is any real number greater than one excluding even and odd integers, the EKZ filter does have limitations to the suppression of certain frequencies. In this case the EKZ filter still strongly suppresses signals with a period shorter than the filter window length, but it no longer completely eliminates signals with a period equal to, or a multiple of, the filter window length. Despite that limitation in this case, the EKZ filter can still be an effective low-pass filter, with a now continuous choice of filter window size, and even though it does not always completely suppress signals with a period equal to, or a multiple of, the filter window length, the level of suppression can be controlled though iteration.

Another advantage of the EKZ filter, is the relative ease of calculation in its iterative form, a feature it shares with KZ and SMA filters. While the EKZ can programmed into statistical software, it is also possible to calculate in most spreadsheet programs. As cited earlier, the KZ filter has seen extensive use in time series for signal separation, component signal reconstruction, multivariate analysis, and forecasting. In some cases, the KZ or even SMA filter may be sufficient to perform these functions. However, the KZ and SMA filters are less precise tools than the EKZ, and there may be applications where the precision and fine adjustment possible with the EKZ may be necessary, such as the real data application examples provided. It is possible to imagine a dataset with two component periodic signals with different but very close frequencies. To separate the two component signals it may be necessary to use a filter to split those frequencies, passing one and attenuating the other. There are scenarios where this may not be possible with KZ filters, but the task can be accomplished through the flexibility of the EKZ.

## 6. Conclusion

In time series analysis, the SMA is often used to smooth the time series and reduce shorter period, higher frequency, variation. The KZ filter extended the simple moving average and provided greater control over the frequencies which are attenuated and those that are allowed to pass through the filter, and how much attenuation is used. However, both SMA and KZ filters do not permit sufficient flexibility in choosing the length of the filter window over which the moving average is applied, preventing precise design of the filters. Since they suppress frequencies of *1/m* and shorter, where *m* is the chosen filter window but is restricted to positive odd integers, at times they provide what can best be described as a blunt tool for this task. The problem is most pronounced for small values of *m*, with shorter period signals and higher frequencies, since the corresponding frequency gaps created by the frequencies *1/m* are substantial. The EKZ filter proposed here extends both the KZ and SMA filters to permit a continuous selection of filter window lengths across the same range. As a result, the frequency specifically chosen for attenuation by the filter is no longer limited to the values of *1/m*, but to any real value greater than one. The EKZ creates a much more precise tool with many potential future applications.